\documentclass[10pt]{article}
\usepackage{color}
\usepackage{graphicx}
\usepackage{url}  
\usepackage{multicol}
\makeatletter

\newenvironment{figurehere}
  {\def\@captype{figure}}
  {}
  
\def\section{\@startsection {section}{1}{\z@}
{-3.0ex plus -1ex minus -.2ex}{0.5ex plus .2ex}{\large\bf}}
\def\subsection{\@startsection {subsection}{1}{\z@}
{-1.5ex plus -1ex minus -.2ex}{0.5ex plus .2ex}{\normalsize\bf}}
\makeatother

\begin{document}
\noindent
{\Large\bf KAGRA: 2.5 Generation Interferometric Gravitational Wave Detector
}\\
{KAGRA collaboration}
\footnote{The list of authors is attached in the end of this article.  \\Contact persons of this article: 
Chunglee Kim, chunglee.kim@ewha.ac.kr; 
Yuta Michimura, michimura@granite.phys.s.u-tokyo.ac.jp; 
Hisaaki Shinkai, hisaaki.shinkai@oit.ac.jp; 
Ayaka Shoda, ayaka.shoda@nao.ac.jp}

\begin{multicols}{2}
{\bf \small The recent detections of gravitational waves\cite{GW150914PRL, GW151226PRL, GW170104PRL, GW170608ApJL, GW170814PRL, GW170817PRL} (GWs) reported by LIGO\cite{aligo}/Virgo\cite{adV} collaborations have made significant impact on physics and astronomy. A global network of GW detectors will play a key role to solve the unknown nature of the sources in coordinated observations with astronomical telescopes and detectors. Here we introduce KAGRA (former name LCGT; Large-scale Cryogenic Gravitational wave Telescope), a new GW detector with two 3-km baseline arms arranged in the shape of an “L”, located inside the Mt.\ Ikenoyama, Kamioka, Gifu, Japan. KAGRA's design is similar to those of the second generations such as Advanced LIGO/Virgo, but it will be operating at the cryogenic temperature with sapphire mirrors. This low temperature feature is advantageous for improving the sensitivity around 100 Hz and is considered as an important feature for the third generation GW detector concept (e.g. Einstein Telescope\cite{ET} of Europe or Cosmic Explorer\cite{CE} of USA). Hence, KAGRA is often called as a 2.5 generation GW detector based on laser interferometry. The installation and commissioning of KAGRA is underway and its cryogenic systems have been successfully tested in May, 2018. KAGRA's first observation run is scheduled in late 2019, aiming to join the third observation run (O3) of the advanced LIGO/Virgo network. In this work, we describe a brief history of KAGRA and highlights of main feature. We also discuss the prospects of GW observation with KAGRA in the era of O3. When operating along with the existing GW detectors, KAGRA will be helpful to locate a GW source more accurately and to determine the source parameters with higher precision, providing information for follow-up observations of a GW trigger candidate. 
}

~\\

\small 

Seeing is believing. We have been reminded of this proverb when we received the news of the discovery of GW150914, the first direct detection of gravitational waves (GWs)   \cite{GW150914PRL}. The existence of GWs has been believed, since Russel Hulse and Joseph Taylor discovered the binary pulsar PSR B1913$+$16 in 1974 \cite{HulseTaylor}.  The long-term radio observation of this system showing that the observed orbital decay is well described by the energy/angular momentum loss due to GW emission as predicted by Einstein in 1915  \cite{TaylorWeisberg}. However, the direct detection of GWs had an extraordinary impact not only to the scientific community but also to the general public. 

The first five GW sources \cite{GW150914PRL,GW151226PRL,GW170104PRL, GW170608ApJL,GW170814PRL} were identified to be binary black holes (BHs). They have mass ranges between $20 \sim 60~M_\odot$, which are heavier than known BHs in our Galaxy as stellar ones. In addition to the confirmation of the existence of binary BHs itself, which is one of the scientific achievements, more GW observations will allow us to understand better about the formation and evolution of binary BHs. 

The latest event GW170817 \cite{GW170817PRL}, was the long-sought event of binary neutron star (NS) merger. The distance and location of GW170817 was narrowed down to $40\pm8$ Mpc and about 30 deg$^2$ in the sky by the LIGO-Virgo observation, allowing astronomers to identify electromagnetic counterparts and the host Galaxy (NGC 4993) \cite{mmaGW170817}. Furthermore, {\it afterglows} from the merger remnants and later outcomes via various baryonic interactions were observed by the telescopes on the Earth as well as space satellites  from radio to $\gamma$-rays.

\begin{center}
\begin{figurehere}
\includegraphics[width=7.5cm,clip]{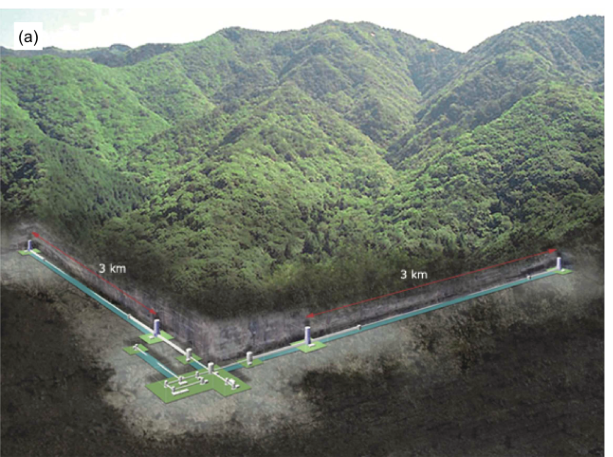} \\
\includegraphics[width=7.5cm,clip]{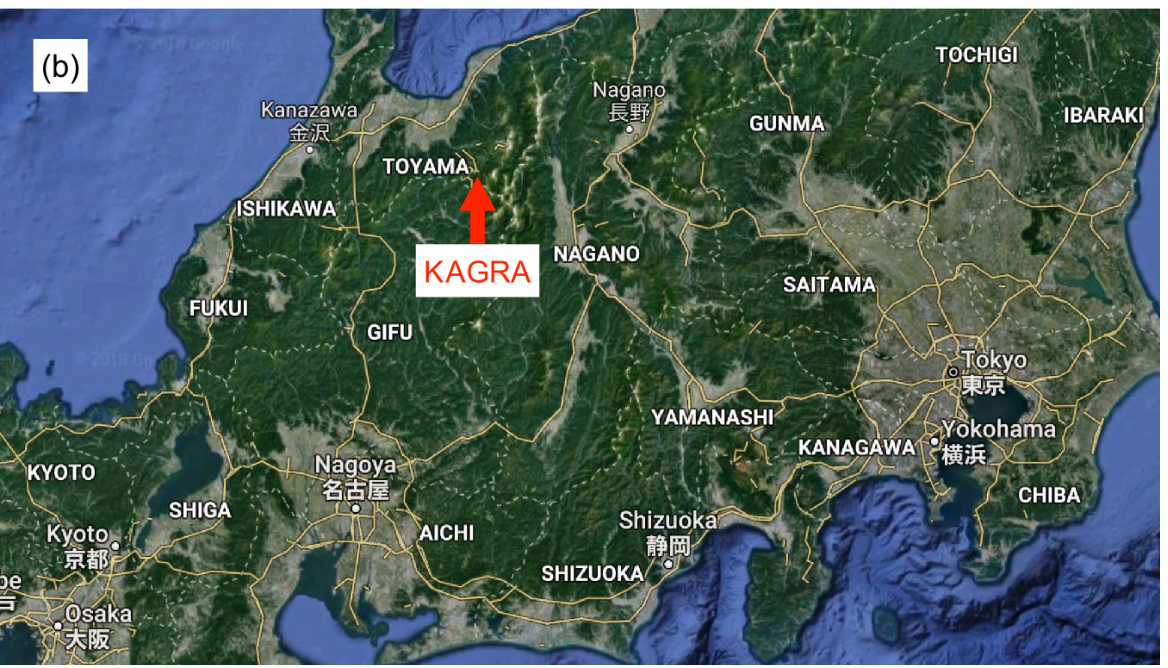}
\caption{\footnotesize LOCATION OF KAGRA. (a) Concept image of KAGRA: a 3-km cryogenic interferometer inside Ikenoyama mountain. (b) KAGRA is located in Kamioka, Gifu, Japan, which is located 220 km northwest of Tokyo. \label{fig1}}
\end{figurehere}
\end{center}

All these discoveries are success stories of the long-baseline laser interferometers as a highly effective  tool to explore the universe via GWs. LIGO consists of two 4-km laser interferometers in Livingston, Louisiana and Hanford, Washington in the USA. Virgo is a 3-km interferometer located in Pisa, Italy. Coincident signal-extraction analyses of these three detectors can eliminate false detections due to noise, and by using triangulation, the source location in the sky can be determined within several tens of square degrees. For a more precise source localization and binary parameter estimation, it is essential to extend the global network of GW detectors, with KAGRA being the next to come online.  

Fig.~\ref{fig1} shows the location of KAGRA, where the interferometer shares the area with the well-known neutrino detectors, {\it Super-Kamiokande} and {\it KamLAND}. Kamioka is a small town, with its biggest claim to fame being an old mine located at 1.5 hour driving distance from the city of Toyama. 

Comparing to existing laser interferometers, KAGRA is technologically unique in two features. First, it is located in an {\it underground} site in order to reduce the seismic noise. In addition, KAGRA's test masses are sapphire mirrors that are designed to be operated at cryogenic temperatures ($\sim $20 K) in order to reduce thermal noise. 
KAGRA is a resonant sideband extraction (RSE) interferometer \cite{rse}, and quantum non-demolition techniques \cite{qnd} are planned to be applied to beat the standard quantum limit of displacement measurements. As a result, KAGRA is expected to reach an equivalent sensitivity to those of Advanced LIGO/Virgo; $2 \times 10^{-24} /{\rm \sqrt{Hz}}$ at 100~Hz.

\section*{Milestones of KAGRA construction and operations}
In Japan, plans to construct interferometric GW detectors started in the 1980s. In the early 90s, the Institute of Space and Astronautical Science (ISAS) and the National Astronomical Observatory of Japan (NAOJ) constructed a 100-m delay-line Michelson interferometer (TENKO-100)  \cite{TENKO100} and a 20-m Fabry-Perot Michelson interferometer \cite{20mproto}, respectively. The former realized 102-times-long light paths of the arm length (the equivalent of a 10.2-km arm length) and reached a sensitivity of $1.1 \times 10^{-19}/\sqrt{{\rm Hz}}$ in the frequency range of 800 Hz -- 2.5 kHz.  

In 1995, the construction of a 300-m Fabry-Perot Michelson interferometer, called TAMA (or TAMA300) \cite{TAMA1997}, began in the Mitaka campus of the NAOJ. The TAMA detector's name is originated from the area where NAOJ is located in. After 3 years of commissioning, the TAMA interferometer was operated for the first time in 1998 with a sensitivity ($5\times 10^{-21}/\sqrt{{\rm Hz}}$). In 2000, the 40-m prototype of LIGO was built, but the 1998 sensitivity achieved by TAMA remained to be unbeaten. 

In 2001, TAMA was successfully operated for more than 1000~hours \cite{TAMA300} and in 2002, it also took part in a joint operation with LIGO's 2nd science run (S2) for two months. TAMA was planned as a prototype in order to develop future technologies for a km-scale interferometer which included a power-recycling system and a seismic attenuation system (SAS). TAMA's final (and best) sensitivity of $1.3 \times 10^{-21} /\sqrt{{\rm Hz}}$ was obtained at around 1~kHz.

TAMA was located in a city of Mitaka, a suburb of Tokyo. In the frequency band below 100~Hz, therefore, significant seismic noise due to human activities in and around the mega city was inevitable. In order to overcome the large seismic noise, it was decided to put a planned future interferometer underground.  An old mine in a mountain in Kamioka was selected as the site of this new interferometer and  experiments for early commissioning began. The LISM \cite{LISM} (Laser Interferometer Small Observatory in a Mine) project (2000--2002) brought a 20-m Fabry-Perot interferometer from NAOJ to Kamioka and confirmed that the Kamioka site is less affected by seismic noise than the Tokyo/Mitaka area. The LISM and TAMA groups performed a simultaneous observation and the first veto analysis  that aims to remove false triggers caused by the instrument \cite{vetoTAMA2004,vetoLIGO2010}.

The Cryogenic Laser Interferometer Observatory (CLIO)~\cite{CLIO} was constructed next in Kamioka from 2002. CLIO was an interferometer with two perpendicular 100-m arms and sapphire mirrors were installed and cooled down to 20~K. The operation started in 2005 and the experiments continued until 2010. This system reduced various thermal noises and the seismic noise was two orders of magnitude lower than that of the Tokyo area.

Although various experiments showcased the  possible scientific achievements of the project and the plausibility of fundamental technologies, the proposal for developing a km-scale cryogenic GW detector was in limbo for many years. This was mainly due to the fact that there was no GW detection reported in 2000s. Without a detection, the proposed km-baseline interferometer concept LCGT~\cite{LCGT} was considered to be too expensive and too risky.
The current was changed when Takaaki Kajita became the director of the institute for cosmic ray research (ICRR, Univ. of Tokyo) in 2008.  He realized the situation and  decided to lead the GW project by his own account, that was the starting point of the current project. 
The LCGT project was finally approved in 2010 with the starting budget 14 billion JPY (150 million USD) for construction, and the excavation of the tunnels in Kamioka began in 2012, after an one-year delay due to the 2011 Tohoku Earthquake. During the construction, LCGT was given its nickname, KAGRA, chosen from a public naming contest. The name KAGRA is taken from KAmioka (the location) plus GRAvity; the Japanese word {\it kagura} reminds a type of traditional sacred dance accompanied by music dedicated to gods.

After a two-year excavation and another two-year facility installation period, KAGRA performed a test operation in March and April 2016 with a simple 3-km Michelson interferometer configuration, called {\it iKAGRA} (initial KAGRA)~\cite{iKAGRA}. 

The strain sensitivity of iKAGRA was limited by seismic noise below 3 Hz, by acoustic noise over 100~Hz to 3~kHz, and by sensor noise at 3--5~kHz. Unfortunately, a series of large earthquakes hit the Kumamoto area during the period of iKAGRA operations. Such noise sources were not avoidable with the iKAGRA configuration, but iKAGRA still provided the collaboration with invaluable experiences in controlling the km-scale laser interferometer with unprecedented sensitivity. 

\begin{center}
\begin{figurehere}

\includegraphics[width=8.25cm,clip]{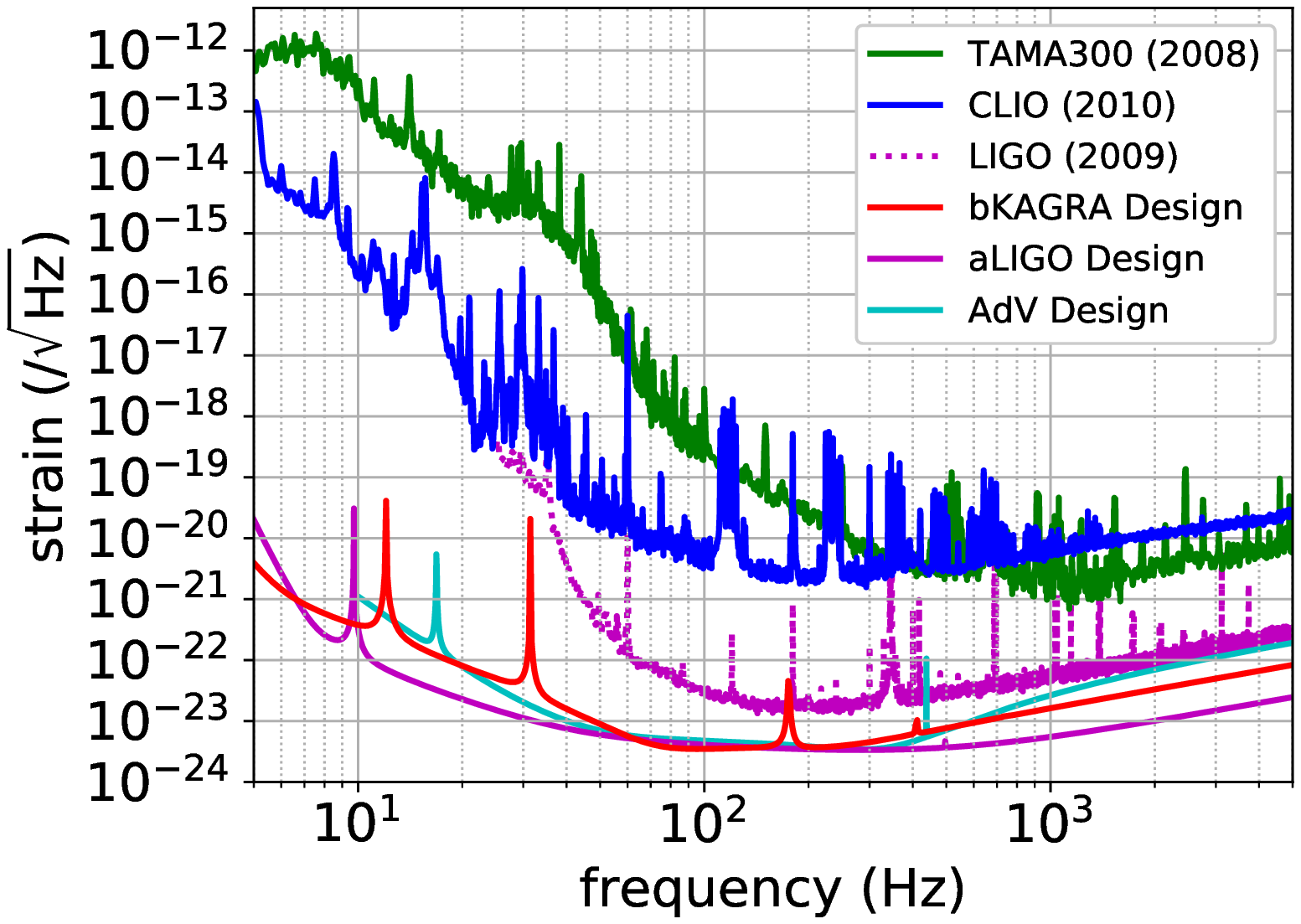}\\

\caption{\footnotesize EVOLUTION OF GW DETECTORS IN JAPAN. We compare the measured sensitivity curves obtained from TAMA300 (2008) and CLIO (2010) with the design sensitivity of the bKAGRA, which is to be accomplished in 2022 in the current plan. The sensitivity curves of LIGO during the science run in 2009-2010 and of Advanced LIGO (aLIGO)/Advanced Virgo (AdV) designs are also shown.\label{fig_TAMAKAGRA}}
\end{figurehere}
\end{center}

\section*{Current status of KAGRA}

Right after the iKAGRA operation, the KAGRA collaboration put great effort and spent for two more years in upgrading of the whole system: installing cryogenic facilities, upgrading vibration-isolation systems and high vacuum systems. This period is called the {\it phase-1} of {\it bKAGRA} (baseline KAGRA). KAGRA is the world's third largest vacuum system; the first two are LIGO-Livingston and LIGO-Handford. 

One of the major upgrades is the vacuum system; KAGRA now has the world's tallest vibration isolation systems (13.5 m) which help to reduce seismic noise at low frequencies. Two 23-kg sapphire mirrors have been installed at both ends, and one of them was kept at 18~K for 30 days continuously.

Due to a leakage of the vacuum that was found in April 2018, the experimental operation was delayed for five days, but the phase-1 operation was successfully undertaken for 9 days from April 28 to May 6, 2018.  
The duty cycle of the first 5 days of the phase-1 operation reached 88.6\% between April 28 and May 2, while the duty cycle on May 3 \& 4 dropped to 26.8\%, and slightly improved to 59.8\% on the final days (May 5 \& 6). The low duty cycle on May 3 and afterwards were mainly attributed to the micro-seismic noise caused by a heavy storm, local earthquakes, and volcano eruptions in Hawaii. The obtained sensitivity during the phase-1 was worse than the final sensitivities of TAMA and CLIO, apart from the lower frequencies, where KAGRA's sensitivity was indeed better than TAMA. (see the designed sensitivity of bKAGRA in Fig.~\ref{fig_TAMAKAGRA}).
More detailed results of the phase-1 operation will be reported elsewhere.

On May 7 2018, the KAGRA collaboration announced the beginning of {\it phase-2} and has been working on the installations/upgrades of more instruments, such as additional optics and a new higher power laser source. 

Since the beginning of the iKAGRA commissioning, one of the important goals of the KAGRA collaboration has been to contribute to the international efforts for GW detection. As shown by the LIGO-Virgo joint observations of GW170814 and GW170817, independent detections at different sites on Earth will be not only useful for increasing the signal-to-noise ratio of a GW trigger signal but also important to better constrain the GW source location. Increasing the number of detectors is also intrinsically essential to distinguish polarization modes of GWs. When KAGRA will be added to the global network of GW detectors, the total network sensitivity will be plausible to search for signatures due to non-tensorial GWs and to more stringently test general relativity \cite{Takeda}.

With this in mind, there have been calls from both within and outside of the collaboration to take part in joint observations. In order to take advantages of a quadrupole detector network around the Earth for GW search, the KAGRA collaboration has been investing great effort to accelerate its commissioning schedule than the plan shown in the ``Scenario paper" \cite{LRR2018}. If everything goes well, KAGRA will join the LIGO/Virgo's 3rd observation run (O3), which is planned to start from February 2019 and last for a year.

The latest road map of KAGRA's GW observation is presented in Fig.~\ref{figLVK}, along with the time line of the LIGO/Virgo observation plans.  This figure is an update of Fig.~2 in \cite{LRR2018}. By March 2019, almost all the optics for the final interferometer configuration are to be installed.  After some tuning of the interferometer, KAGRA plans to begin its observation phase no later than October 2019.  If RSE implementation is ready, then KAGRA will start with 25~Mpc observational range in NS-NS binary coalescence, which distance is in the same level of Virgo in the end of O2.
While waiting the operation of the interferometer ready, the data analysis groups in KAGRA are planning to start co-data analysis with LIGO/Virgo groups from the beginning of O3. After O3, KAGRA will also try to catch up with LIGO/Virgo's O4 schedule from its beginning.
 The main interferometer components will be planned to be installed by March 2019, and for O4, we expect to approach the designed sensitivity at the horizon distance of 130~Mpc for NS-NS binary inspirals in the broadband configuration.

\begin{center}
\begin{figurehere}
\includegraphics[width=7.5cm,clip]{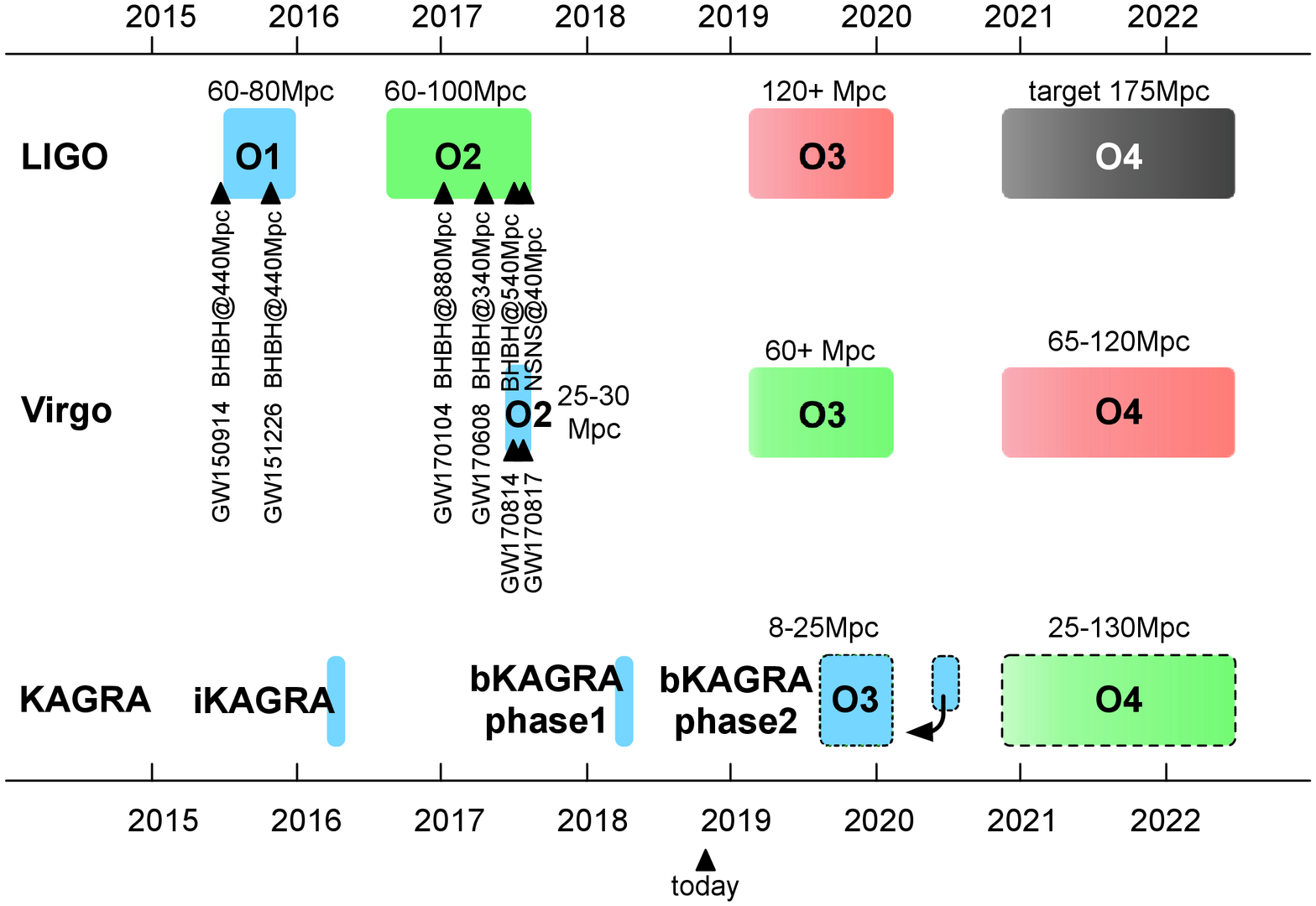}
\caption{\footnotesize OBSERVATION PLANS OF LIGO, VIRGO, AND KAGRA. An un-offical update of Fig.~2 presented in \cite{LRR2018}. The color bars indicate the first (blue), second (green), and third  (red) observation period. Details of the KAGRA's long-term plan is still in discussion. In principle, the collaboration aims that, after joining O3 of LIGO/Virgo in its later phase, KAGRA will begin an observation run from the beginning of O4 coordinated with other detectors. 
\label{figLVK}}
\end{figurehere}
\end{center}

\section*{KAGRA international collaboration}
The KAGRA collaboration is by all means international. As of October, 2018, the collaboration consists of more than 200 researchers from 90 institutions in 15 countries and regions. KAGRA collaboration has a decision making body in science, named KAGRA Scientific Congress (KSC). KSC organizes the KAGRA international workshops outside of Japan twice every year and interact with a broader scientific community.  

KSC is also setting the organization for future joint observations, and is now re-organizing data-analysis groups in order to match their structures with those of LIGO/Virgo groups.  Data analysis teams are preparing the original analysis codes, called ``KAGRA Algorithmic LIbrary" (KAGALI); {\it kagari} is also the Japanese word, bonfire at the celebration party. The code includes, for example, a couple of new methods for extracting ring-down waveform solely with a purpose for testing gravity theories, and such new ideas are expected to enhance the advantages of joint observations with LIGO/Virgo.

Including the ELiTES project in 2012-2017, KAGRA collaboration has been trying to expand international collaboration. For example,  the ``core-to-core (C2C)'' program funded by Japan Society for the Promotion of Science (JSPS) is dedicated to human resource exchanges and organizing academic meetings relevant to KAGRA and GW science. 

In addition to local computing centers in Kamioka observatory, ICRR, and Osaka City Univ., etc., KAGRA collaboration runs two mirror sites for data storage in Korea (KISTI) and in Taiwan (Academia Sinica). KISTI and Academia Sinica also provide computing resources for data analysis.

\section*{KAGRA's future}
The LIGO and Virgo collaborations have plans to upgrade their detectors to A+ \cite{a+} and AdV+ \cite{adv+} to improve the sensitivity by a factor of roughly 2 compared with the original Advanced LIGO and Advanced Virgo designs. For example, both A+ and AdV+ detectors will incorporate frequency dependent squeezing to reduce quantum noise and lower-loss coating to reduce coating thermal noise.  

Similarly, the KAGRA collaboration have recently started  planning for an upgrade of KAGRA to realize the binary neutron-star range of more than 150~Mpc. KAGRA has a unique potential to improve the sensitivity because of its cryogenic operation and lower seismic noise compared to LIGO and Virgo. The completion of the A+ and AdV+ upgrades is expected by $\sim 2023$, and the KAGRA collaboration aims for a similar timeline for KAGRA.

Cryogenic operation and underground construction are expected to be key technologies of the third generation large-scale GW detectors. For example, collaborative R\&D activities between KAGRA and the next generation projects, as represented by the ELiTES (ET-LCGT Telescopes: Exchange of Scientists) project have been supported by the Japanese government and European Commission in 2012-2017.

{\small

~\\
The KAGRA project could be possible to be realized backed by enormous support from the scientific community as well as a great effort by researchers around the world.The upgraded {\it bKAGRA} is about to see the ``first light" in 2019. Nobody doubts the importance of GW research in astronomy, physics, and also in engineering. We believe KAGRA will definitely contribute to these fields, especially to GW science, and help to broaden our understanding of gravity and of the Universe.

\subsection*{Acknowledgment}
This work was supported by MEXT, JSPS Leading-edge Research Infrastructure Program, JSPS Grant-in-Aid for Specially Promoted Research 26000005, JSPS Grant-in-Aid for Scientific Research on Innovative Areas 2905: JP17H06358, JP17H06361 and JP17H06364, JSPS Core-to-Core Program A. Advanced Research Networks, JSPS Grant-in-Aid for Scientific Research (S) 17H06133, the joint research program of the Institute for Cosmic Ray Research, University of Tokyo, National Research Foundation (NRF) grant of Korea and Computing Infrastructure Project of KISTI-GSDC in Korea, the LIGO project, and the Virgo project. 
The authors appreciate Marios Karouzos for his kind feedback to improve the draft. 
}
\end{multicols}

\noindent
{\bf KAGRA collaboration}\\
\noindent
\small
T. Akutsu$^{1, 2}$, 
M. Ando$^{3, 4, 1}$, 
K. Arai$^{5}$, 
Y. Arai$^{5}$, 
S. Araki$^{6}$, 
A. Araya$^{7}$, 
N. Aritomi$^{3}$, 
H. Asada$^{8}$, 
Y. Aso$^{9, 10}$, 
S. Atsuta$^{11}$
% ...
\\

\noindent \footnotesize
${}^{1}$ National Astronomical Observatory of Japan (NAOJ), Mitaka City, Tokyo 181-8588, Japan\\
${}^{2}$ Advanced Technology Center, National Astronomical Observatory of Japan (NAOJ), Mitaka City, Tokyo 181-8588, Japan\\
${}^{3}$ Department of Physics, The University of Tokyo, Bunkyo-ku, Tokyo 113-0033, Japan\\
${}^{4}$ Research Center for the Early Universe (RESCEU), The University of Tokyo, Bunkyo-ku, Tokyo 113-0033, Japan\\
${}^{5}$ Institute for Cosmic Ray Research (ICRR), KAGRA Observatory, The University of Tokyo, Kashiwa City, Chiba 277-8582, Japan\\
${}^{6}$ Accelerator Laboratory, High Energy Accelerator Research Organization (KEK), Tsukuba City, Ibaraki 305-0801, Japan\\
${}^{7}$ Earthquake Research Institute, The University of Tokyo, Bunkyo-ku, Tokyo 113-0032, Japan\\
${}^{8}$ Department of Mathematics and Physics, Hirosaki University, Hirosaki City, Aomori 036-8561, Japan\\
${}^{9}$ Kamioka Branch, National Astronomical Observatory of Japan (NAOJ), Kamioka-cho, Hida City, Gifu 506-1205, Japan\\
${}^{10}$ The Graduate University for Advanced Studies (SOKENDAI), Mitaka City, Tokyo 181-8588, Japan\\
${}^{11}$ Graduate School of Science and Technology, Tokyo Institute of Technology, Meguro-ku, Tokyo 152-8551, Japan\\
${}^{12}$ Institute for Cosmic Ray Research (ICRR), KAGRA Observatory, The University of Tokyo, Kamioka-cho, Hida City, Gifu 506-1205, Japan\\
${}^{13}$ Korea Institute of Science and Technology Information (KISTI), Yuseong-gu, Daejeon 34141, Korea\\
${}^{14}$ Graduate School of Science, Osaka University, Toyonaka City, Osaka 560-0043, Japan\\
${}^{15}$ Department of Physics, National Taiwan Normal University, Taipei 116, Taiwan R.O.C.\\
${}^{16}$ Department of Physics, Sogang University, Mapo-Gu, Seoul 121-742, Korea \\
${}^{17}$ Institute of Physics, Academia Sinica, Nankang, Taipei 11529, Taiwan, R.O.C. \\
${}^{18}$ Department of Physics, University of Toyama, Toyama City, Toyama 930-8555, Japan\\
${}^{19}$ Univ. Grenoble Alpes, Laboratoire d'Annecy de Physique des Particules (LAPP), Universit\'e Savoie Mont Blanc, CNRS/IN2P3, F-74941 Annecy, France\\
${}^{20}$ Department of Astronomy, The University of Tokyo, Bunkyo-ku, Tokyo 113-0033, Japan\\
${}^{21}$ Department of Applied Physics, Fukuoka University, Jonan, Fukuoka City, Fukuoka 814-0180, Japan\\
${}^{22}$ Faculty of Engineering, University of Toyama, Toyama City, Toyama 930-8555, Japan\\
${}^{23}$ Institute for Cosmic Ray Research (ICRR), Research Center for Cosmic Neutrinos (RCCN), The University of Tokyo, Kashiwa City, Chiba 277-8582, Japan\\
${}^{24}$ Yukawa Institute for Theoretical Physics (YITP), Kyoto University, Sakyou-ku, Kyoto City, Kyoto 606-8502, Japan\\
${}^{25}$ Graduate School of Science, Osaka City University, Sumiyoshi-ku, Osaka City, Osaka 558-8585, Japan\\
${}^{26}$ JAXA Institute of Space and Astronautical Science, Chuo-ku, Sagamihara City, Kanagawa 252-0222, Japan\\
${}^{27}$ Institute for Cosmic Ray Research (ICRR), The University of Tokyo, Kashiwa City, Chiba 277-8582, Japan\\
${}^{28}$ Department of Physics, Ewha Womans University, Seodaemun-gu, Seoul 03760, Korea\\
${}^{29}$ Department of Physics, Myongji University, Yongin 449-728, Korea\\
${}^{30}$ Department of Computer Simulation, Inje University, Gimhae, Gyeongsangnam-do 50834, Korea\\
${}^{31}$ National Institute for Mathematical Sciences, Daejeon 34047, Korea\\
${}^{32}$ School of Natural Science, Ulsan National Institute of Science and Technology (UNIST), Ulsan 44919, Korea\\
${}^{33}$ Applied Research Laboratory, High Energy Accelerator Research Organization (KEK), Tsukuba City, Ibaraki 305-0801, Japan\\
${}^{34}$ Department of Physical Science, Hiroshima University, Higashihiroshima City, Hiroshima 903-0213, Japan\\
${}^{35}$ Department of Physics and Institute of Astronomy, National Tsing Hua University, Kuang-Fu Road, Hsinchu 30013, Taiwan R.O.C.\\
${}^{36}$ Institute for Cosmic Ray Research (ICRR), Research Center for Cosmic Neutrinos (RCCN), The University of Tokyo, Kamioka-cho, Hida City, Gifu 506-1205, Japan\\
${}^{37}$ California Institute of Technology,  Pasadena, CA 91125, USA\\
${}^{38}$ Institute for Advanced Research, Nagoya University, Chikusa-ku, Nagoya City, Aichi 464-8602, Japan\\
${}^{39}$ Department of Physics, Hanyang University, Seoul 133-791, Korea\\
${}^{40}$ Korea Astronomy and Space Science Institute (KASI), Daejeon 34055, Korea\\
${}^{41}$ National Applied Research Laboratories, Hsinchu Science Park, Hsinchu City 30076, Taiwan R.O.C. \\
${}^{42}$ Department of Physics, Tamkang University, Danshui Dist., New Taipei City 25137, Taiwan, R.O.C. \\
${}^{43}$ Department of Advanced Materials Science, The University of Tokyo, Kashiwa City, Chiba 277-8582, Japan\\
${}^{44}$ Istituto Nazionale di Fisica Nucleare, Sapienza University, Roma 00185, Italy\\
${}^{45}$ The Institute of Statistical Mathematics, Tachikawa City, Tokyo 190-8562, Japan\\
${}^{46}$ School of Physics, Korea Institute for Advanced Study (KIAS), Seoul 02455, Korea\\
${}^{47}$ Institute for Photon Science and Technology, The University of Tokyo, Bunkyo-ku, Tokyo 113-8656, Japan\\
${}^{48}$ Disaster Prevention Research Institute, Kyoto University, Gokasho, Uji City, Kyoto 611-0011, Japan\\
${}^{49}$ Institute for Laser Science, University of Electro-Communications, Chofu City, Tokyo 182-8585, Japan\\
${}^{50}$ National Institute of Information and Communications Technology (NICT), Koganei City, Tokyo 184-8795, Japan\\
${}^{51}$ Department of Physics, Kyoto University, Sakyou-ku, Kyoto City, Kyoto 606-8502, Japan\\
${}^{52}$ Faculty of Law, Ryukoku University, Fushimi-ku, Kyoto City, Kyoto 612-8577, Japan\\
${}^{53}$ Department of Physics , University of Notre Dame, Notre Dame, IN 46556, USA\\
${}^{54}$ Department of Physics, Wuhan Institute of Physics and Mathematics, CAS, Xiaohongshan, Wuhan 430071, China\\
${}^{55}$ The University of Shanghai for Science and Technology, Shanghai 200093, China\\
${}^{56}$ Faculty of Engineering, Niigata University, Nishi-ku, Niigata City, Niigata 950-2181, Japan\\
${}^{57}$ Graduate School of Science and Technology, Niigata University, Nishi-ku, Niigata City, Niigata 950-2181, Japan\\
${}^{58}$ Center for Measurement Standards,  Industrial Technology Research Institute, Hsinchu, 30011, Taiwan, R.O.C.\\
${}^{59}$ Department of Engineering, University of Sannio, Benevento 82100, Italy\\
${}^{60}$ Faculty of Arts and Science, Kyushu University, Nishi-ku, Fukuoka City, Fukuoka 819-0395, Japan\\
${}^{61}$ Research Institute for Science and Engineering, Waseda University, Shinjuku, Tokyo 169-8555, Japan\\
${}^{62}$ National Institute of Technology, Nagaoka College, Nagaoka City, Niigata 940-8532, Japan\\
${}^{63}$ Kavli Institute for the Physics and Mathematics of the Universe (IPMU), Kashiwa City, Chiba 277-8583, Japan\\
${}^{64}$ Nagaoka University of Technology, Nagaoka City, Niigata 940-2188, Japan \\
${}^{65}$ Graduate School of Science and Engineering, Hosei University, Koganei City, Tokyo 184-8584, Japan\\
${}^{66}$ Faculty of Science, Toho University, Funabashi City, Chiba 274-8510, Japan\\
${}^{67}$ Faculty of Information Science and Technology, Osaka Institute of Technology, Hirakata City, Osaka 573-0196, Japan\\
${}^{68}$ School of High Energy Accelerator Science, The Graduate University for Advanced Studies (SOKENDAI), Tsukuba City, Ibaraki 305-0801, Japan\\
${}^{69}$ Istituto Nazionale di Fisica Nucleare, University of Perugia, Perugia 06123, Italy\\
${}^{70}$ Faculty of Science, Niigata University, Nishi-ku, Niigata City, Niigata 950-2181, Japan\\
${}^{71}$ Department of Communications, National Defense Academy of Japan, Yokosuka City, Kanagawa 239-8686, Japan\\
${}^{72}$ Department of Physics, University of Florida, Gainesville, FL 32611, USA\\
${}^{73}$ Department of Physics and Astronomy, Sejong University, Gwangjin-gu, Seoul 143-747, Korea \\
${}^{74}$ Department of Physics, Korea University, Seongbuk-gu, Seoul 02841, Korea\\
${}^{75}$ Department of Astronomy, Beijing Normal University, Beijing 100875, China

\begin{thebibliography}{99}

\bibitem{GW150914PRL} Abbott, B. P. et al. Observation of Gravitational Waves from a Binary Black Hole Merger. {\it Phys. Rev. Lett.} {\bf 116}, 061102 (2016).

\bibitem{GW151226PRL} Abbott, B. P. et al. GW151226: Observation of Gravitational Waves from a 22-Solar-Mass Binary Black Hole Coalescence. {\it Phys. Rev. Lett.} {\bf 116}, 241103 (2016).

\bibitem{GW170104PRL} Abbott, B. P. et al. GW170104: Observation of a 50-Solar-Mass Binary Black Hole Coalescence at Redshift 0.2. {\it Phys. Rev. Lett.} {\bf 118}, 221101 (2017).

\bibitem{GW170608ApJL} Abbott, B. P. et al. GW170608: Observation of a 19 Solar-mass Binary Black Hole Coalescence, {\it Astrophys. J. Lett.} {\bf 851}, L35 (2017).

\bibitem{GW170814PRL} Abbott, B. P. et al. GW170814: A Three-detector Observation of Gravitational Waves from a Binary Black Hole Coalescence . {\it Phys. Rev. Lett.} {\bf 119}, 141101 (2017).

\bibitem{GW170817PRL} Abbott, B. P. et al. GW170817: Observation of Gravitational Waves from a Binary Neutron Star Inspiral. {\it Phys. Rev. Lett.} {\bf 119}, 161101 (2017).

\bibitem{aligo} Aasi, J. et al. Advanced LIGO, {\it Class. Quant. Grav.} {\bf 32}, 074001 (2015).

\bibitem{adV} Acernese, F. et al. Advanced Virgo: a second-generation interferometric gravitational wave detector. {\it Class. Quant. Grav.} {\bf 32}, 024001 (2015).

\bibitem{ET} Punturo, M. et al. The Einstein Telescope: a third-generation gravitational wave observatory. {\it Class. Quant. Grav.} {\bf 27}, 194002 (2010).

\bibitem{CE} Abbott, B. P. et al. (the LIGO Collaboration) Exploring the sensitivity of next generation gravitational wave detectors. {\it Class. Quant. Grav.} {\bf 34}, 044001 (2017). 

\bibitem{HulseTaylor} Hulse, R. A. \& Taylor, J. H. Discovery of a pulsar in a binary system. {\it Astrophys. J.} {\bf 195}, L51 (1975). 
\bibitem{TaylorWeisberg} Taylor, J. H. \& Weisberg, J. M. Gravitational radiation and the binary pulsar. {\it Astrophys. J.}  {\bf 253}, 908 (1981).

\bibitem{mmaGW170817} Abbott, B. P. et al. Multi-Messenger Observations of a Binary Neutron Star Merger. {\it Astrophys. J. Lett.} {\bf 848}, L12 (2017).

\bibitem{rse} Mizuno, J. et al. Resonant sideband extraction: a new configuration for interferometric gravitational wave detectors. {\it Phys. Lett. A} {\bf 175}, 273 (1993).

\bibitem{qnd} Somiya, K. (KAGRA Collaboration) Detector configuration of KAGRA - the Japanese cryogenic gravitational-wave detector. {\it Class. Quant. Grav.} {\bf 29}, 124007 (2012).

\bibitem{TENKO100} Kawashima K., Laser Interferometer (TENKO-10 and -100) for Gravitational Wave Antenna Development, The Institute of Space and Astronautical Science Report, 640 (1991).

\bibitem{20mproto} Sato S. et al. High-gain power recycling of a Fabry?Perot Michelson interferometer for a gravitational-wave antenna, {\it Applied Optics} {\bf 36}, 1446 (1997).

\bibitem{TAMA1997} Tsubono K., in {\it Gravitational Wave Experiments}, edited by Coccia E., Pizzella G., and Ronga F., (World Scientific, 1995), p. 112-114.

\bibitem{TAMA300} Ando, M. et al. Stable Operation of a 300-m Laser Interferometer with Sufficient Sensitivity to Detect Gravitational-Wave Events within Our Galaxy. {\it Phys. Rev. Lett.} {\bf 86}, 3950 (2001).

\bibitem{LISM} Sato, S. et al. Ultrastable performance of an underground-based laser interferometer observatory for gravitational waves. {\it Phys. Rev. D} {\bf 69}, 102005 (2014).

\bibitem{vetoTAMA2004} Ando, M. et al. (the KAGRA collaboration). Analysis methods for burst gravitational waves with TAMA data. {\it Class. Quant. Grav.} {\bf 21}, S1679 (2004).

\bibitem{vetoLIGO2010} Christensen, N. et al. (LIGO Collaboration, Virgo Collaboration), {\it Class. Quant. Grav.} {\bf 27}, 194010 (2010).

\bibitem{CLIO} Uchiyama, T. et al. Reduction of Thermal Fluctuations in a Cryogenic Laser Interferometric Gravitational Wave Detector. {\it Phys. Rev. Lett.} {\bf 108}, 141101 (2012).

\bibitem{LCGT} 
Kuroda, K. et al. Large-Scale Cryogenic Gravitational Wave Telescope, {\it Int. J.  Mod. Phys. D} {\bf 8}, 557 (1999).

\bibitem{iKAGRA} Akutsu, T. et. al. Construction of KAGRA: an underground gravitational-wave observatory. {\it Prog. Theor. Exp. Phys.} {\bf 2018}, 013F01 (2018).

\bibitem{Takeda} Takeda, H. et al. Polarization test of gravitational waves from compact binary coalescences. {\it Phys. Rev. D} {\bf 98}, 022008 (2018).

\bibitem{LRR2018} Abbot, B. P. et al. Prospects for observing and localizing gravitational-wave transients with Advanced LIGO, Advanced Virgo and KAGRA. {\it Liv. Rev. Rel.} {\bf 21}, 3 (2018).

\bibitem{a+} Miller, J. et al. Prospects for doubling the range of Advanced LIGO {\it Phys. Rev. D} {\bf 91}, 062005 (2015).

\bibitem{adv+} Degallaix, J. for the Virgo Collaboration. Advanced Virgo+ preliminary studies, The Virgo Collaboration, VIR-0300A-18 (2018).
\end{thebibliography}
\end{document}